\documentclass[journal=jctcce,manuscript=article]{achemso}

\usepackage[version=3]{mhchem} 
\usepackage[T1]{fontenc}
\usepackage{color}
\usepackage{textcomp}
\usepackage[nolist]{acronym}
\usepackage{fixltx2e}
\usepackage{amsmath}
\usepackage{leftidx}

\definecolor{background-color}{gray}{0.98}

\author{Julius P. P. Zauleck and Regina de Vivie-Riedle}
\affiliation{Department Chemie, Ludwig-Maximilians-Universit\"at M\"unchen, D-81377 M\"unchen, Germany}
\email{regina.de_vivie@cup.uni-muenchen.de}

\title{Constructing grids for molecular quantum dynamics using an autoencoder}


\begin{document}

%
%
%

\newpage

\begin{figure}[h]
\centering
\colorbox{background-color}{
\fbox{
\begin{minipage}{1.0\textwidth}
\includegraphics[width=110mm,height=50.6mm]{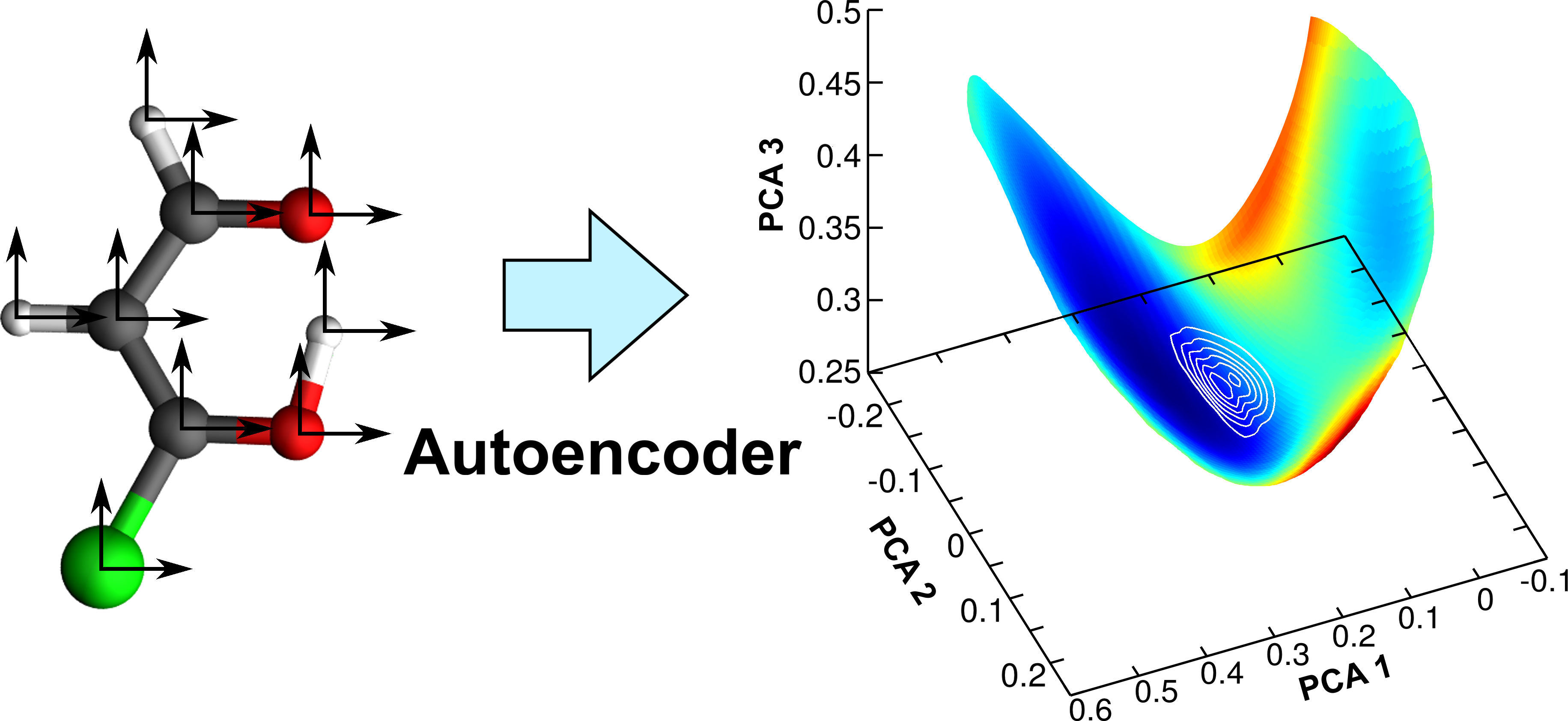} 
\\
The large number of degrees of freedom of typical systems make molecular quantum dynamics very challenging. One approach is to reduce the dimensionality of the system and construct a grid in that space on which the quantum dynamics calculations are performed. Finding this generally highly unintuitive nonlinear subspace is a difficult task. In this paper we introduce a machine learning based approach using lower level trajectory calculations as training data.
\end{minipage}
}}
\end{figure}
\begin{acronym}
\acrodef{pca}[PCA]{principal component analysis}
\acrodef{coin}[CoIn]{conical intersection}
\acrodef{irc}[IRC]{intrinsic reaction coordinate}
\acrodef{ts}[TS]{transition state}
\acrodef{dft}[DFT]{density functional theory}
\acrodef{pes}[PES]{potential energy surface}
\acrodef{si}[SI]{supporting information}
\acrodef{iba}[IBA]{IRC based approach}
\acrodef{tba}[TBA]{trajectory based approach}
\acrodef{qd}[QD]{quantum dynamics}
\acrodef{rbm}[RBM]{restricted Boltzmann machine}
\end{acronym}

\begin{abstract}
A challenge for molecular quantum dynamics (QD) calculations is the curse of dimensionality with respect to the nuclear degrees of freedom. A common approach that works especially well for fast reactive processes is to reduce the dimensionality of the system to a few most relevant coordinates. Identifying these can become a very difficult task, since they often are highly unintuitive. We present a machine learning approach that utilizes an autoencoder that is trained to find a low-dimensional representation of a set of molecular configurations. These configurations are generated by trajectory calculations performed on the reactive molecular systems of interest. The resulting low-dimensional representation can be used to generate a potential energy surface grid in the desired subspace. Using the G-matrix formalism to calculate the kinetic energy operator, QD calculations can be carried out on this grid. In addition to step-by-step instructions for the grid construction, we present the application to a test system.

\end{abstract}
%

\section*{Introduction}
The search for an efficient basis probably remains as the central challenge in molecular wavepacket \ac{qd}. Several different approaches have been developed, each with its most efficient field of application. They all, in one way or another, have to deal with the exponential growth of their basis and therefore scaling of computation time with the number of degrees of freedom. The probably most successful approach, the multi-configuration time-dependent Hartree method \cite{Meyer89,Meyer2003}, finds a powerful basis while requiring the potential part of the Hamiltonian to take on a certain form. This is highly efficient for a many systems but in cases where the wavefunction moves far away from potential minima, like in reactive scenarios, grid based approaches that are restricted to fewer degrees of freedom can be more practical. In this case, choosing the degrees of freedom of a subspace that best describe the reaction is one of the most important steps to produce reliable predictions. While we recently presented two ways to automate the search for linear subspaces \cite{Zauleck2016}, the identification of even more powerful nonlinear subspaces is still to a large degree driven by intuition. In this work we present an approach to automate this identification based on the data produced by trajectory calculations, replacing human intuition by a machine learning algorithm.

The search for nonlinear subspaces is generally much more difficult than for linear ones. For them a mathematical tool exists to extract the dimensions of biggest variance of a data set, namely the principal component analysis. For a nonlinear generalization, the shape of the subspace now enters as an additional optimization task and there is no exact mathematical solution. Nevertheless, there are a number of approximate solutions that have gained popularity in recent years and many of them have been applied to molecular problems.\cite{Brown2008,Roweis2000,Tenenbaum2000,Das2006,Nguyen2006,Manuscript2012,Amadei1999,Mansbach2016} These methods can all be summarized with the concept of nonlinear dimensionality reduction. Due to its demonstrated efficiency for molecular data \cite{Brown2008} as well as the promising prospect of further development \cite{Vincent2010,Sankaran2017}, we chose to use an autoencoder, a type of neural network, for our problem.

Neural networks have been extensively applied to the direct calculation of \ac{pes}s in the last few decades and have signficantly improved the system sizes that can be described.\cite{Handley2014,Blank1995,Manzhos2006b,Manzhos2006,Behler2007,Houlding2007,Pukri2009,Handley2010,Behler2011,Nguyen2012,Jiang2013,Behler2014,Gastegger2015,Gastegger2016} In our case, we use an autoencoder to learn a low-dimensional representation of a reaction, in which we then calculate a \ac{pes}. For this, an autoencoder takes high-dimensional input in the form of data points corresponding to molecular geometries and finds a low-dimensional embedding that describes them well. We can then use this embedding to extract reactive coordinates and to span a grid, on which the time-dependent Schrödinger equation can be solved. Using trajectories is one way to generate input data points to train the autoencoder. Although these will generally miss the quantum effects we wish to include, they still can give meaningful information about the parts of molecular configuration space used to describe the reaction.

Once the grid in the desired subspace is created, the \ac{pes} can be calculated for each grid point, completing the potential part of the Hamiltonian. Constructing the kinetic energy operator is more complicated, but the G-matrix formalism \cite{Wilsonbook} that has proven useful for \ac{qd} in reactive coordinates \cite{Stare2003,Kowalewski2014,Thallmair2016} can be applied here, too.

In the following section the construction, training and use of the autoencoder describing the subspace will be explained in detail. This includes the construction of a training data set from a chosen set of trajectories and a step-by-step explanation how to construct a grid that is ready for \ac{qd} simulations. For completeness, we also give a short overview over the G-matrix formalism. Thereafter an application to the example system of the proton transfer in (Z)-hydroxyacryloyl chloride is presented. Our approach will be explained in detail using the example of a pseudo-spectral grid representation, making use of the fast Fourier transform on rectangular grids.\cite{Kosloff1983} However, it is generally applicable for any scenario where a set of nonlinear coordinates is needed and a data set of trajectory points can be created that is approximately contained in those coordinates. 

\section*{Methodology}
There are several approaches to find a nonlinear subspace in the context of molecular dynamics. For grid based \ac{qd} this is often referred to as the search for reactive coordinates and we will refer to the nonlinear subspace in which the \ac{qd} is performed as reactive coordinates throughout this work. While there exist a few approaches like Z-matrix coordinates or Jacobi coordinates that are straightforward and describe some chemical reactions relatively well using very few coordinates, often linear combinations of them are more efficient.\cite{Luckhaus2000,Luckhaus2004,Hofmann2000} Finding those combinations is mostly driven by experience and intuition and even then, there are many cases where the reactive coordinates best suited are simply of a more general kind. This behavior can already be observed in the \ac{irc} for a chemical transformation. The \ac{irc} rarely consist of a linear combination of Z-matrix or Jacobi Coordinates and arguably are very close to the best one-dimensional subspace to describe a reaction.

As a result, our goal is to find a generalization that can recreate the aforementioned special cases of reactive coordinates as well as the \ac{irc}, but that is also not limited to any specific form. We will present such an approach using an autoencoder. In the molecular context we usually think of an reactive coordinate as a change of relative orientations of atoms. As we search for a generalized nonlinear subspace, these relations between the atoms become less intuitive, because we do not bias the algorithms towards our intuitions. While the autoencoder will generate a number of 'raw' reactive coordinates that span the subspace, they will usually be overly nonlinear and thus not very useful for understanding molecular motions or creating a \ac{pes} grid. However, we will be left with a mapping between the subspace and the full-dimensional configuration space and vice versa. These mappings will then allow us to project configuration points onto the subspace and in turn give us the ability to create a \ac{pes} spanned by more useful reactive coordinates.

In the next section, we will introduce the setup and functionality of an autoencoder in some detail to ensure comprehensibility for readers that are less familiar with the topic.

\subsection*{Autoencoder}
The growth of computer performance and the development of new algorithms have caused an explosion in the applicability of machine learning methods. Probably, the recently most influential methods come from the field of artificial neural networks and an important type of neural networks are autoencoders. 

\begin{figure} [ht]
\begin{center}
   \includegraphics[width=0.6\columnwidth,keepaspectratio=true]{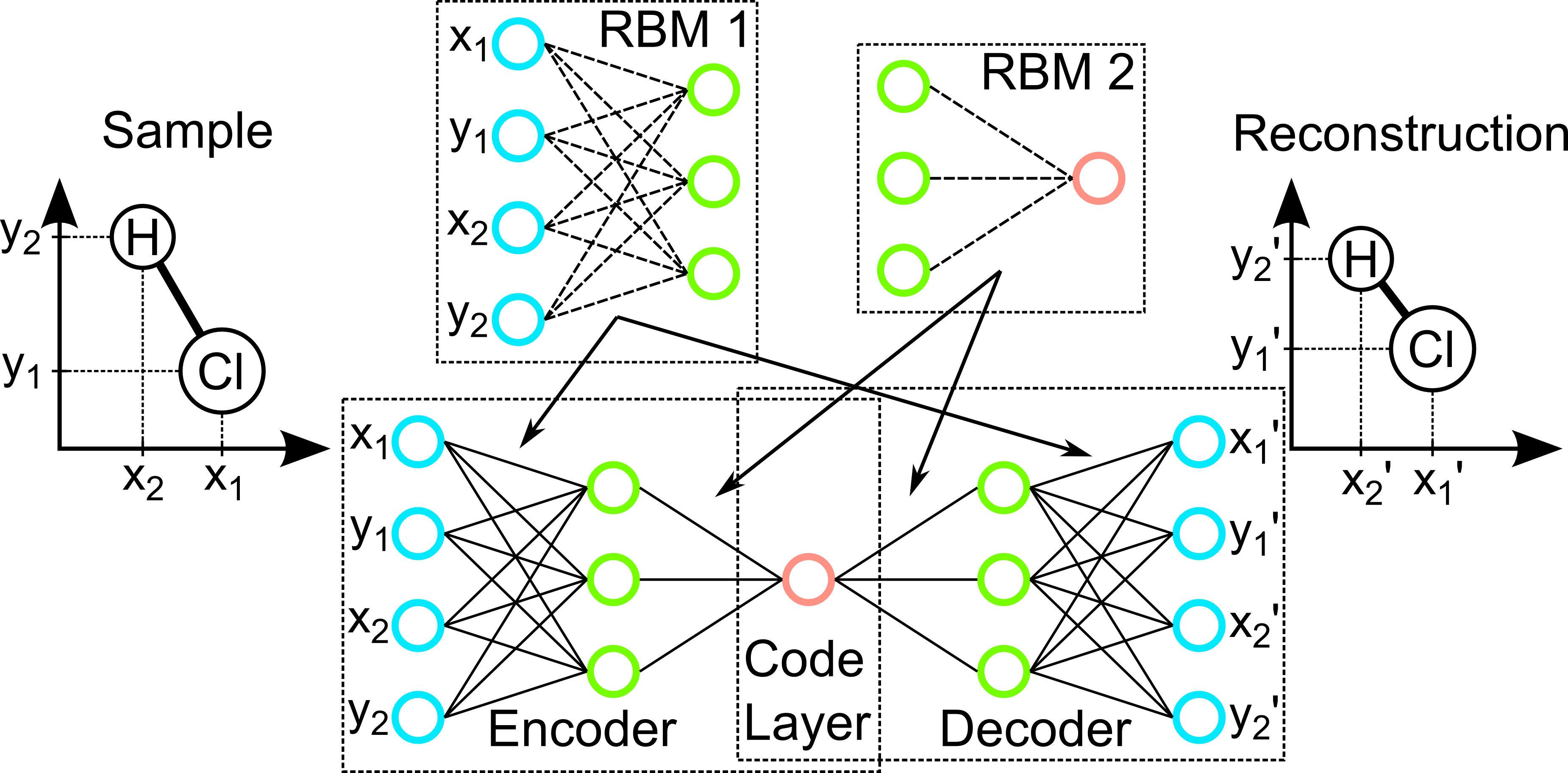} 
\end{center}
\vspace{0.25in}
\hspace*{3in}
  \caption{Schematic of an autoencoder reproducing molecular configurations of hydrochloric acid. Each Cartesian coordinate of each atom is represented by an input node. The \ac{rbm}s architectures are chosen according to the autoencoder's. Their weights --here shown as dashed connections-- can then be used as initial weights for the autoencoder.}
  \label{fig:auto}
\end{figure}

\ref{fig:auto} illustrates the setup of an autoencoder on the example of molecular configurations. If each Cartesian coordinate of the shown molecule is represented by an input node and the autoencoder architecture bottleneck is made of a single node, one can use the autoencoder to reproduce molecular geometries according to their interatomic distances. Here we refer to the number of layers of the network with their respective numbers of nodes as the architecture. In our case, autoencoders consist of an encoder and a decoder, where the encoder has an input layer of nodes, zero or more hidden layers (green nodes in \ref{fig:auto}) and a layer of output nodes that represent the code (red). The decoder takes this code layer as input, has the same number of hidden layers with the same numbers of nodes as the encoder in opposite order and has an output layer with the same number of nodes as the input layer of the encoder. With respect to the complete autoencoder, coming from the input layer, all nodes of a following layer $i$ starting with $i=1$ take the function values $\mathbf{z}^{i-1}$ of the nodes of the previous layer $i-1$ as variables $\mathbf{x}^{i}$ to calculate their function values $\mathbf{z}^{i}$:
\begin{align}
\mathbf{z}^{i}=\sigma^i\left(\mathbf{W}^i\mathbf{x}^i+\mathbf{b}^i\right) \text{ with } \mathbf{x}^i=\mathbf{z}^{i-1}
\label{eq:autoenc1}
\end{align}
These in turn become the input arguments of the next layer. In the example of \ref{fig:auto} $\mathbf{x}^{1}$ corresponds to the Cartesian coordinates of the sample geometries and $\mathbf{z}^{4}$ corresponds to the Cartesian coordinates of the reconstructed geometries. Given an input vector with $P$ dimensions $\mathbf{W}^i$ is a weight matrix of dimensionality $Q\times P$ and $\mathbf{b}^i$ a bias vector of dimensionality $P$, producing an output vector with $Q$ dimensions. $\sigma^i(\mathbf{u})$ is an element-wise activation function of $\mathbf{u}$ introducing non-linearity between the layers. In our examples $\sigma^i$ will always be the logistic function $\sigma^i_l(\mathbf{u})=1/\left(1+\exp(\mathbf{u}_l)\right)$ except for the calculation of the output layer of the encoder, i.e. the code layer, for which it will be the identity function. Here the $l$th element of the output vector is calculated from the $l$th element of the input vector. The autoencoder can be trained by providing a sufficiently large number of input samples and correcting weights and biases so the input samples are better reproduced.

However, for practical autoencoders with many hidden layers training can be difficult when the initial weights and biases are not close to a good solution. It has been demonstrated that finding such initial weights can be done by layer-by-layer pretraining using a \ac{rbm}.\cite{Hinton2006} A \ac{rbm} consists of a visible and a hidden layer and learns a probability distribution over its possible inputs (in the visible layer) by calculating gradients of its weights and biases in order to increase the probabilities of the input samples. These weights and biases can be initialized with normal distributions and zero values, respectively. After training the set of activations of the hidden layer corresponding to the set of input samples they can be used to train the next \ac{rbm}. If the architectures of the \ac{rbm}s are chosen according to the architecture of the autoencoder, the weights and biases of the \ac{rbm}s can be used as the initial weights and biases of the respective layers of the encoder and decoder. After this, the complete autoencoder can be trained by calculating the gradients of the weights and biases in order to produce more accurate results using the chain rule, which is known as backpropagation.\cite{Williams1986} This is equivalent to minimizing the mean reconstruction error $\frac{1}{N}\sum_{k}^N\left|\left|\mathbf{x}^1_k-\mathbf{z}^{M+1}_k\right|\right|$ given $N$ training samples, $M$ hidden layers and superscript $1$ representing the input layer. For the example given in \ref{fig:auto} this corresponds to the distance vectors between the sampled and reconstructed Cartesian coordinates of the atoms. The training steps in machine learning are usually counted in epochs, where one epoch refers to a machine training over the whole set of training data once.

Once successfully trained, the final weights and biases are determined and the autoencoder can approximately reproduce input data in its output layer. This allows us to choose a bottleneck architecture where the code layer of the autoencoder only has very few nodes, equalling the number of degrees of freedom aimed for. In this case, the autoencoder learns an approximate reconstruction of the learning data using the number of variables equal to the number of nodes in the code layer. We can use this approximate reconstruction to create an approximate projection of full-dimensional data points onto the low-dimensional subspace defined by the code layer and then refine this projection iteratively using the decoder part of the autoencoder.

To obtain the refined projection we can start with two mappings between the full-dimensional space and the subspace best describing the training data points. These are similar to the encoder and decoder, but behave optimally in a way that will be described in the following. The first mapping $\mathbf{y}'=\mathbf{F}(\mathbf{y})$ takes a full-dimensional vector $\mathbf{y}$ as input and returns a vector $\mathbf{y}'$ with the dimensionality of the subspace as output. The second, inverse mapping $\mathbf{y}''=\mathbf{H}(\mathbf{y}')$ takes such a vector of subspace dimensionality as input and returns a full-dimensional vector $\mathbf{y}''$. Let these two mappings with $\mathbf{y}''=\mathbf{H}(\mathbf{F}(\mathbf{y}))$ have optimal behavior in the sense that if a full-dimensional vector $\mathbf{y}$ lies within the subspace, then $\mathbf{y}''=\mathbf{y}$ and that if $\mathbf{y}$ does not lie within the subspace, the distance $\left|\left| \mathbf{y}'' -\mathbf{y} \right|\right|$ is the smallest possible for the subspace. We can then use the encoder $\widetilde{\mathbf F}$ and the decoder $\widetilde{\mathbf H}$ as approximations of $\mathbf F$ and $\mathbf H$, respectively.

To use them for a refined projection, we apply the autoencoder first to produce an approximate projection $\widetilde{\mathbf{y}}''=\widetilde{\mathbf{H}}(\widetilde{\mathbf{y}}')=\widetilde{\mathbf{H}}(\widetilde{\mathbf{F}}(\mathbf{y}))$. Then, we construct a set of basis vectors $\mathbf{a}_j$ by changing the activation of every individual node $j$ of the coding layer starting from $\widetilde{\mathbf{y}}'$:
\begin{align}
\mathbf{a}_j= \widetilde{\mathbf{H}}\left(\widetilde{\mathbf{y}}'+\epsilon\mathbf{e}_j\right) - \widetilde{\mathbf{H}}\left(\widetilde{\mathbf{y}}'\right)
\label{eq:autoenc2}
\end{align}
Here, $\mathbf{e}_j$ is the $j$th standard basis vector and $\epsilon$ determines the length of the vector. $\epsilon$ should be chosen small to ensure a local basis with respect to $\widetilde{\mathbf{y}}'$ using the decoder. Since the initial projection is approximate, the distance $\left|\left| \widetilde{\mathbf{H}}\left(\widetilde{\mathbf{y}}'\right) -\mathbf{y} \right|\right|$ has generally not reached a local minimum regarding variations in $\widetilde{\mathbf{y}}'$. If the corresponding distance vector is projected onto the basis vectors $\mathbf{a}_j$, this minimum can be found iteratively. Using the non-orthonormal projection matrix $\mathbf{A}=[\mathbf a_1\ \mathbf a_2\,\cdots\,\mathbf a_S]$ with $S$ being the number of dimensions of the coding subspace, the first step in the subspace is given by
\begin{align}
\widetilde{\mathbf{y}}_1'= \widetilde{\mathbf{y}}'+h\epsilon\left(\mathbf{A}^T\mathbf{A}\right)^{-1}\left( \mathbf{y}-\widetilde{\mathbf{H}}\left(\widetilde{\mathbf{y}}'\right)  \right)^T\mathbf{A} 
\label{eq:autoenc3}
\end{align}
and the next steps are given by
\begin{align}
\widetilde{\mathbf{y}}_k'= \widetilde{\mathbf{y}}_{k-1}'+h\epsilon\left(\mathbf{A}^T\mathbf{A}\right)^{-1}\left( \mathbf{y}-\widetilde{\mathbf{H}}\left(\widetilde{\mathbf{y}}_{k-1}'\right)  \right)^T\mathbf{A} \text{ .}
\label{eq:autoenc4}
\end{align}
The step size $h$ could be set to 1 to progress through the full projection, but since the procedure is iterative, the subspace is nonlinear and the projection is approximate, a smaller $h$ is suggested. A derivation of eqs. \ref{eq:autoenc3} and \ref{eq:autoenc4} can be found in the Supporting Information.

\subsection*{Training Data Generation}
In order to learn the representation of a subspace, the autoencoder ideally is trained on equally distributed data points over the subspace region that will be used for \ac{qd} calculations. There are many example systems and therefore many different ways to produce the training data set. One general way is to use trajectories to simulate the same process as is going to be simulated using \ac{qd} and we will exclusively focus on trajectory based approaches.

For this to work as intended, two criteria must be met. First, the lack of the quantum effects should not have such an impact that the trajectories explore a very different subspace. This is related to a very common problem for grid based wavepacket \ac{qd}. Since the movement of the wavepacket is not known beforehand, a guess has to be made in order to create a grid. In general, trajectories are a good basis for such a guess. It is also possible to incorporate other information of the path, the wavepacket will follow, as we will demonstrate with our example system. Second, the data set should not have very disjoint reaction pathways spanned by different degrees of freedom for the reactions studied. This would result in regions of the subspace between those pathways that would have to be described without available learning data. However, scenarios where this happens are generally costly for grid based \ac{qd} due to the increase in required dimensions and the distinct pathways could in such cases  be studied individually. As a result, this generally does not limit the applicability of our approach.

We chose to present a way to create a subspace for reactive \ac{qd} using trajectories started from the \ac{irc}. There are two reasons for this decision. First, this will create an excellent test scenario for our approach and second, the \ac{irc} is one of the most easily available sources of information about a system. A reasonable requirement for the subspace is to capture the curvature of the \ac{irc} and to additionally include degrees of freedom with a shallow \ac{pes}, in order to allow the wavepacket the most room to move and diverge. Note that in the case of additional critical points one wishes to include, additional trajectory starting points could be included. Staying with the \ac{irc} starting points will also provide an important test how well the autoencoder will solve this task. In general, the number of degrees of freedom of a molecule with a shallow \ac{pes} is still bigger than the number of reactive coordinates we wish to extract. Therefore all additional degrees of freedom with a shallow \ac{pes} that are discarded provide notable noise, because the trajectories will explore them nearly as much as the degrees of freedom that can be included. This will provide an inherent test of the noise resistance of the autoencoder.

\begin{figure} [ht]
\begin{center}
   \includegraphics[width=0.6\columnwidth,keepaspectratio=true]{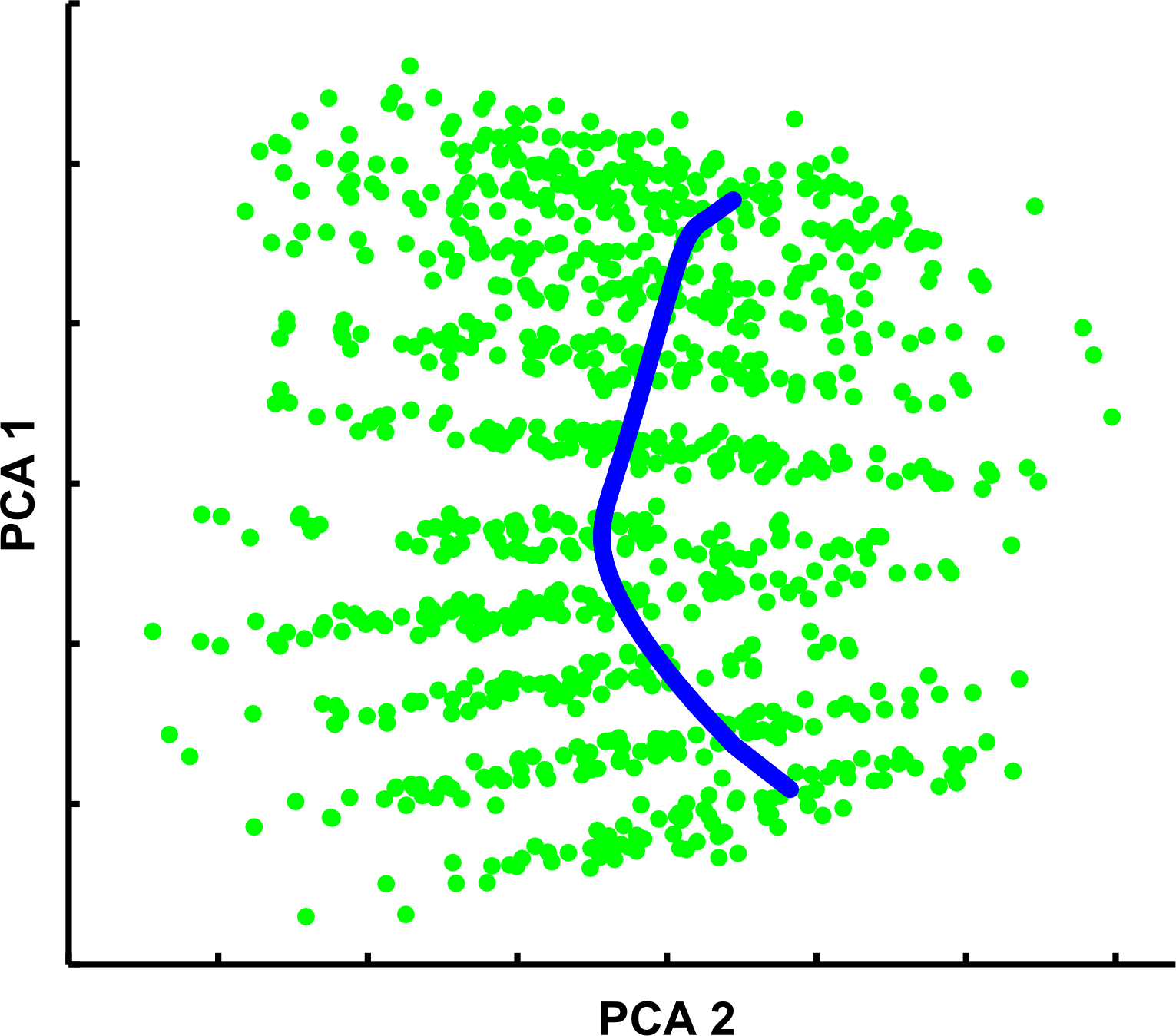} 
\end{center}
\vspace{0.25in}
\hspace*{3in}
  \caption{\ac{irc} (blue) and training data points (green) projected onto the first two principal components. Even though the \ac{irc} is curved in more than three dimensions, the frozen movement of the trajectories parallel to the \ac{irc} at their starting point is clearly visible. Every separate slice of points represents trajectory points from their central \ac{irc} starting point. The topmost points are in this projection not separable anymore, because the \ac{irc} has larger components orthogonal to the first two principal components.}
  \label{fig:datacut}
\end{figure}

To generate data points that enable us to extract the desired subspace, we can not simply let trajectories run along the \ac{irc}, since they will follow its descend and not explore the space orthogonal to it. By incorporating a constraint algorithm like RATTLE \cite{Andersen1983} the movement along a certain direction can be restricted. Now, the data points can be produced by choosing equispaced points along the \ac{irc} and running a number of trajectories from each point while freezing the motion along the \ac{irc}. Since these constraints are of a fairly general kind, they can introduce spurious rotations and translations, which must be avoided by enforcing the Eckart conditions \cite{Eckart1935,Wilsonbook} for the positions and velocities of the atoms at every time step with reference to the starting point of the trajectory. For the trajectories to explore the configuration space efficiently, they can be randomly initialized with momentum orthogonal to the \ac{irc}. Trajectories along degrees of freedom with shallow \ac{pes} will run farther and will be better reconstructed by the autoencoder. We suggest a fixed kinetic energy about the size of the activation energy of the reaction. This will help to ensure the construction of a binding \ac{pes} orthogonal to the \ac{irc}. \ref{fig:datacut} shows a projection of trajectory data points -- corresponding to different geometries -- produced by the described procedure together with the \ac{irc}. For visualization we projected the points on the first two principal components created by a \ac{pca} of the trajectory data points.

\begin{figure} [ht]
\begin{center}
   \includegraphics[width=0.6\columnwidth,keepaspectratio=true]{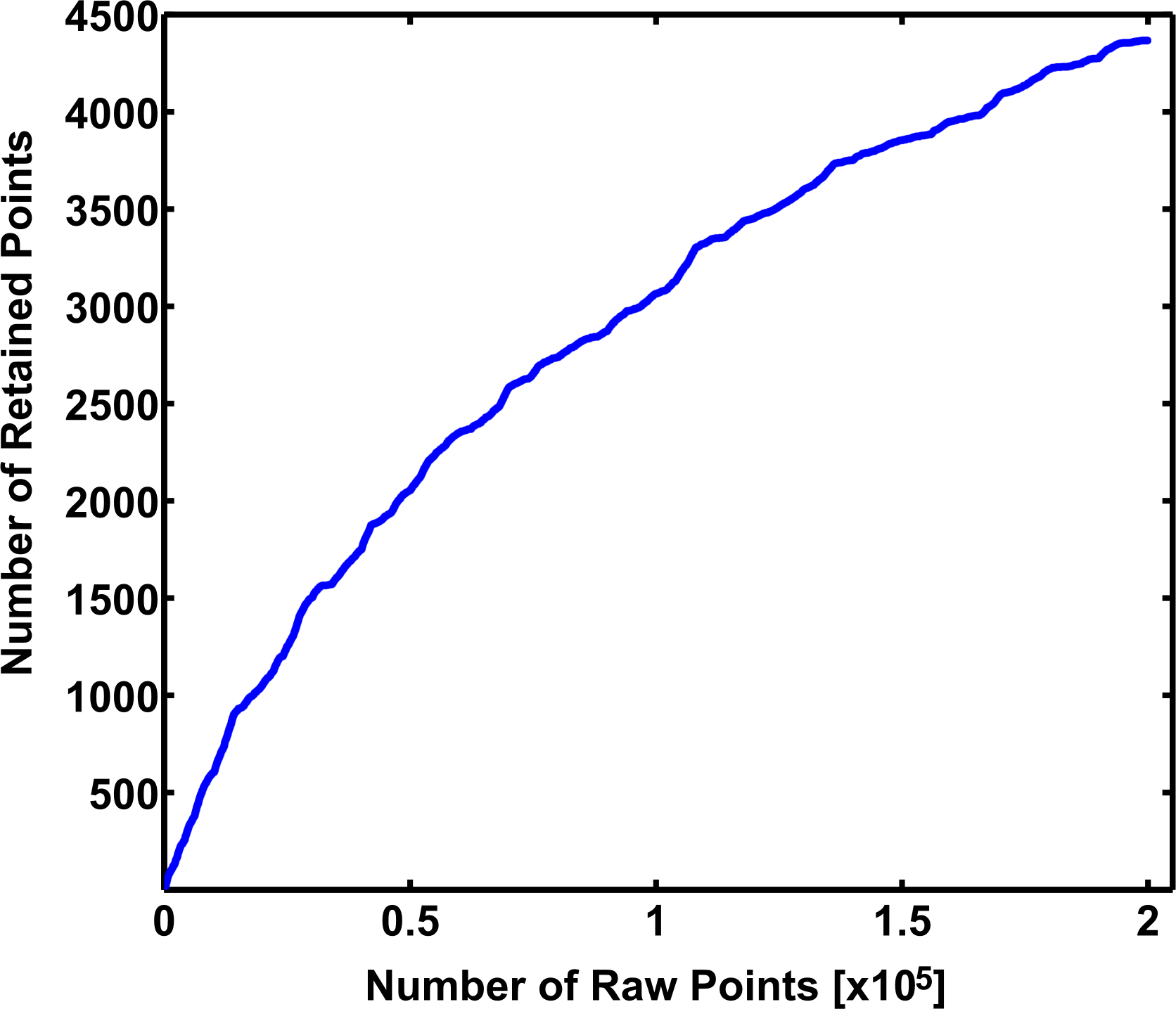} 
\end{center}
\vspace{0.25in}
\hspace*{3in}
  \caption{Convergence behavior of the number of retained data points for a single \ac{irc} starting point. All the 2000 data points of each of the 100 trajectories started from a single \ac{irc} point constitute the raw points. A point qualified to be retained for the training data set, when it was not closer than a certain threshold to any other retained point. Since there is only a limited configuration space to explore with a fixed amount of energy, the number of retained points converges.}
  \label{fig:retain}
\end{figure}

Since we ideally want an equispaced data set, some points of the trajectories starting from the same starting point have to be discarded. This can be done by defining a minimum full-dimensional Euclidean distance between the data points generated and removing all data points that are closer than this distance to any preceding data point. We can use the number of retained points as a function of the number of screened data points to check for convergence. After many trajectory calculations it would be expected that there are only very few new points that are far enough away from all previous points. An example of this trend is shown in \ref{fig:retain}.

All data points from every \ac{irc} starting point are then rescaled so that all components of every input vector lie between 0 and 1. In principal, this data set would be sufficient to train the autoencoder. However, in addition to the pretraining using \ac{rbm}s it is useful and simple to create a pretraining data set with which to train the \ac{rbm}s and the autoencoder first. Thereafter we proceed to training the autoencoder on the actual data set. The pretraining data set can be generated by performing a \ac{pca} on the actual data set. The number of principal components used equals the number of degrees of freedom of the desired subspace. For each principal component the Cartesian coordinate with the biggest contribution is chosen. These now span a linear Cartesian subspace of the same dimensionality as the desired subspace for the \ac{qd} calculation and a random distribution of samples with equal variance along all of those Cartesian coordinates is created. These geometry samples thus only vary within the selected Cartesian coordinates. Since the variance along all degrees of freedom is identical and since only a minority of input nodes are required, this is significantly easier to learn. It is also a good starting point, since an overly curved subspace can lead to unphysical results in the \ac{qd} simulations. As we move to the actual data set during training, additional atomic coordinates and nonlinearities will be introduced until the autoencoder converges.

\subsection*{Grid Construction and G-Matrix}
Once we have successfully trained the autoencoder, we have all information about the subspace, but we still have to construct a practical grid. There are several possible approaches to do this. A general strategy is to choose a subspace in which a rectangular grid is easily constructed and that is similar to the subspace found by the autoencoder (see below). The points of the rectangular grid of the simple subspace can then be projected onto the autoencoder subspace. Since we want to use the G-matrix formalism for the kinetic energy operator, we have to identify the reactive coordinates of the projected grid. Here, we will first give a short introduction to the G-matrix formalism while a more extensive presentation can be found in the literature\cite{Wilsonbook,Schaad1989,Stare2003} and thereafter we introduce the grid construction.

Using the G-matrix $\mathbf{G}$, the kinetic energy operator $\hat T_\mathbf{q}$ in reactive coordinates $\mathbf{q}$ is given by
\begin{align}
  \hat T_\mathbf{q} &= -\frac{\hbar^2}{2} \sum_{r=1}^{N_{\mathbf{q}}} \sum_{s=1}^{N_{\mathbf{q}}} \frac{\partial}{\partial q_r}\left[ \mathbf{G}_{rs} \frac{\partial}{\partial q_s} \right] \text{ .}
\label{eq:gmat1}
\end{align}
$\mathbf{G}$ can be calculated via its inverse, whose elements $\left(\mathbf{G}^{-1}\right)_{rs}$ are given by
\begin{align}
 \left(\mathbf{G}^{-1}\right)_{rs} = \sum_{i=1}^{3N_{\text{at}}} m_i \frac{\partial x_i}{\partial q_r} \frac{\partial x_i}{\partial q_s} \text{ .}
\label{eq:gmat2}
\end{align}
The sums in eq. \ref{eq:gmat1} run over the number of reactive coordinates that are included in the \ac{qd} calculation $N_{\mathbf{q}}$ and the sum in eq. \ref{eq:gmat2} runs over the number of Cartesian coordinates of all atoms $N_{\text{at}}$. As the masses of the atoms $m_i$ are included in it, the G-matrix can be considered a position dependent reciprocal mass that introduces the necessary coupling between the reactive coordinates. Eq. \ref{eq:gmat1} is valid under the assumption that the Jacobian determinant of the coordinate transformation is coordinate independent. As a result it is good practice to keep the volume change of the reactive coordinates approximately constant with respect to the Cartesian coordinates. The projected grids generally fulfil this reasonably well. Besides that, the reactive coordinates the G-matrix can use are very general and as long as we can identify any consistent relation between the change in reactive coordinates and the ordering of the grid points, we can calculate the kinetic energy operator. Since we are focusing on rectangular grids, we can simply assign an reactive coordinate $q_r$ for each direction of our grid and choose arbitrary bounds, e.g. $[0,1]$.

To construct a rectangular grid to be projected, the \ac{pca} can be used again. The principal components of the trajectory training data set can be identified as the directions of a rectangular grid in a linear subspace. The spacing between the grid points should be chosen according to the maximum kinetic energy that has to be represented along all directions on the grid. Using the G-matrix of the \ac{pca} based grid, the spacing $\Delta q_r$ along the different directions $q_r$ should be
\begin{align}
 \Delta q_r < \pi \sqrt{\frac{G_{rr}}{2T_r}}\text{ .}
\label{eq:gmat3}
\end{align}
Here, $T_r$ is the kinetic energy along the direction of the $r$th coordinate that has to be representable. A derivation of eq. \ref{eq:gmat3} can be found in the Supporting Information. In practice, the activation energy gives a good idea of the maximum kinetic energy along the reaction coordinate and some value like half or less of the minimum required spacing is chosen. The number of grid points in each direction is then decided by the critical points that have to be included and by the extension of the \ac{pes} necessary to prevent the wave function from reaching the edges of the grid.

Once the \ac{pca} based rectangular grid is determined, the points can be projected onto the subspace using the autoencoder $\widetilde{\mathbf{H}}(\widetilde{\mathbf{F}}(\mathbf{y}))$ and eqs. \ref{eq:autoenc3} and \ref{eq:autoenc4}. Since every grid point corresponds to a molecular configuration, the \ac{pes} can be calculated using quantum chemistry methods and $\mathbf{G}$ can be calculated via finite differences using adjacent grid points and eq. \ref{eq:gmat2}. The starting wavepacket can be created as in any other grid based calculation depending mostly on the studied reaction.

\section*{Example System: (Z)-hydroxyacryloyl Chloride}
\begin{figure} [ht]
\begin{center}
   \includegraphics[width=0.6\columnwidth,keepaspectratio=true]{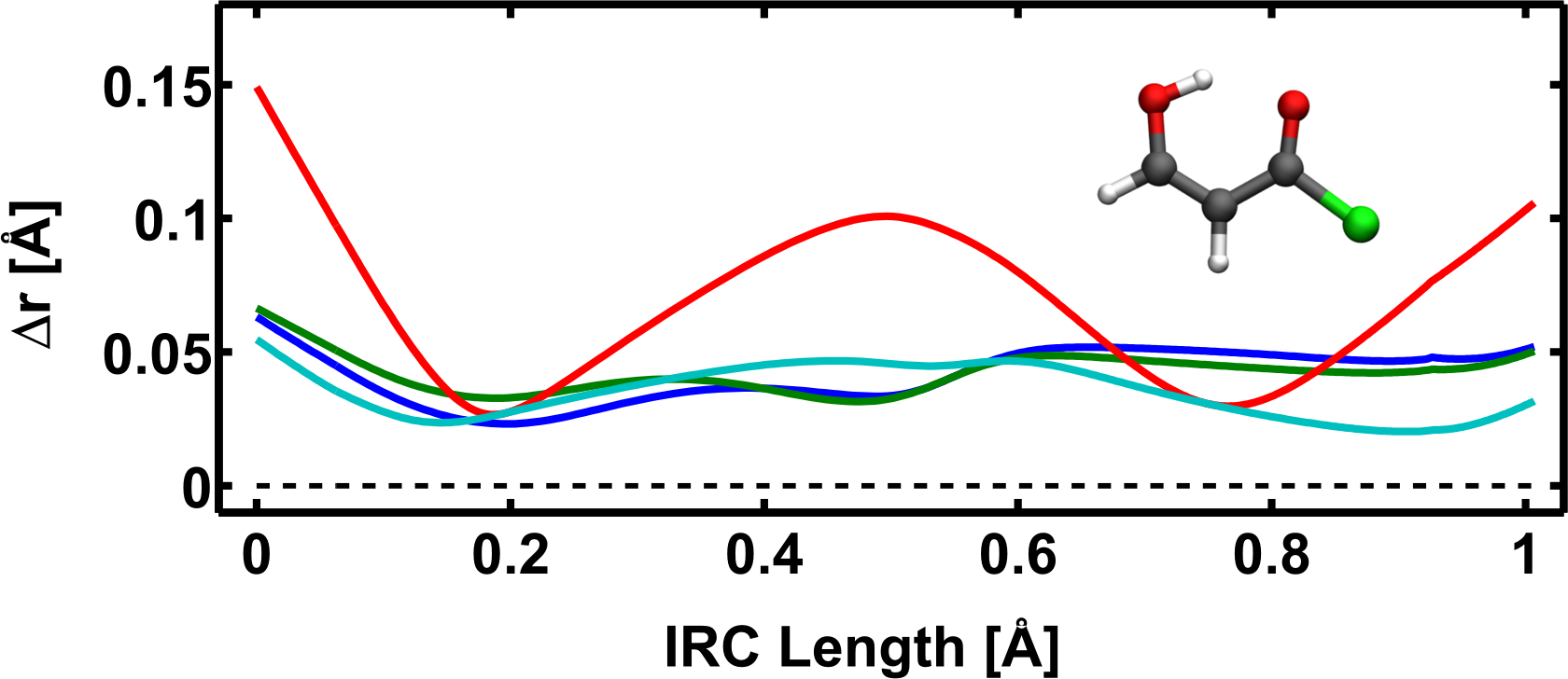} 
\end{center}
\vspace{0.25in}
\hspace*{3in}
  \caption{Distances between projected \ac{irc}s and the full-dimensional one. The blue and green curves represent projections on 2- and 3-dimensional nonlinear subspaces, respectively. These were created by training the autoencoder on the trajectory data sets. The red and cyan curves represent projections on 2- and 3-dimensional linear subspaces, respectively. These were created using the first two or three principal components of the same data set used for the autoencoders. (Z)-hydroxyacryloyl chloride is shown in the upper right.}
  \label{fig:ircdist}
\end{figure}

We tested our method on the proton transfer reaction in (Z)-hydroxyacryloyl chloride. This is an asymmetric variant of the proton transfer in malondialdehyde \cite{Matanovic2008}, where one of the terminal carbon atoms' hydrogen is substituted with chlorine (\ref{fig:ircdist}). The protonization of the oxygen on the far side of the chlorine is much more stable. Since this system requires about two to three degrees of freedom to be appropriately described \cite{Zauleck2016}, we created a two-dimensional and a three-dimensional grid on which \ac{qd} calculations were performed. The number of nodes per layer of the autoencoder used were 18-25-15-7-2 and 18-25-15-7-3. There are 18 input nodes because the model system is planar and nearly stays so during the reaction and has 9 atoms, resulting in 18 degrees of freedom. The data set was created by calculating the \ac{irc} with the program package Gaussian09 \cite{Gaussian09} at the DFT (M06-2X,6-31G(d)) level of theory and picking 10 equispaced points. From each point 100 trajectories were started with 2000 time steps of 0.5 fs using the program package NEWTON-X 1.4.\cite{newtonx}, where the gradients were calculated at HF/6-31G(d) level. This was sufficient for the trajectories since they only have to explore the configuration space. Each trajectory was initialized with a randomly directed momentum corresponding to a kinetic energy of 0.345 eV, which is approximately equal to the activation energy of the reaction. All movement parallel to the \ac{irc} was removed using RATTLE. These two million data points resulted in 38564 training samples after enforcing a minimum distance of 0.087 {\AA}. The two-dimensional and three-dimensional pretraining data sets consisted of 2166 and 9291 training samples, respectively. For each of the three training data sets, a set consisting of 2000 independent test samples was used.

The \ac{rbm}s and autoencoders were trained using data in batches of 2000 samples from the three training sets and the number of epochs of training were 50 for the \ac{rbm}s, 2000 for the autoencoder learning the pretraining data set and 1000 for the autoencoder learning the actual data set. After the 2000 epochs on the pretraining data set, the two-dimensional and three-dimensional autoencoders reproduced the data points with an average error of 0.1056 {\AA} and 0.1319 {\AA}, respectively. The reconstruction of the actual data points after 1000 epochs led to an error of 0.1590 {\AA} and 0.1511 {\AA}, respectively. However, projecting the data points using eqs. \ref{eq:autoenc3} and \ref{eq:autoenc4} lead to more precise reconstructions as can be seen by the deviation of the projected \ac{irc}s from the full-dimensional one shown in \ref{fig:ircdist}. 

Since we used trajectory data points to construct these two autoencoder subspaces, it would not be expected that any low-dimensional projection would reconstruct the \ac{irc} exactly. However, we would expect that, as long as an increase in dimensionality is very beneficial for the description, the distance of projected \ac{irc}s would decrease significantly. We see this decrease between the projections onto 2-dimensional (red) and 3-dimensional (cyan) linear subspaces in \ref{fig:ircdist}. The distances of the nonlinear projections (blue and green) do not differ significantly from the projection onto the 3-dimensional linear subspace. A likely explanation is that two nonlinear or three linear coordinates capture most relevant motions. After this, the description only slowly improves with increasing dimensionality. The most important result is the decrease of distance between the two-dimensional linear subspace (red) and the two-dimensional nonlinear subspace (blue). This clearly shows that our method finds a much better subspace for the reconstruction of the \ac{irc}. Given the lack of significant further decrease with increase of dimensionality, this nonlinear two-dimensional subspace might even capture all of the significant motions of the reaction.

\begin{figure} [ht]
\begin{center}
   \includegraphics[width=0.6\columnwidth,keepaspectratio=true]{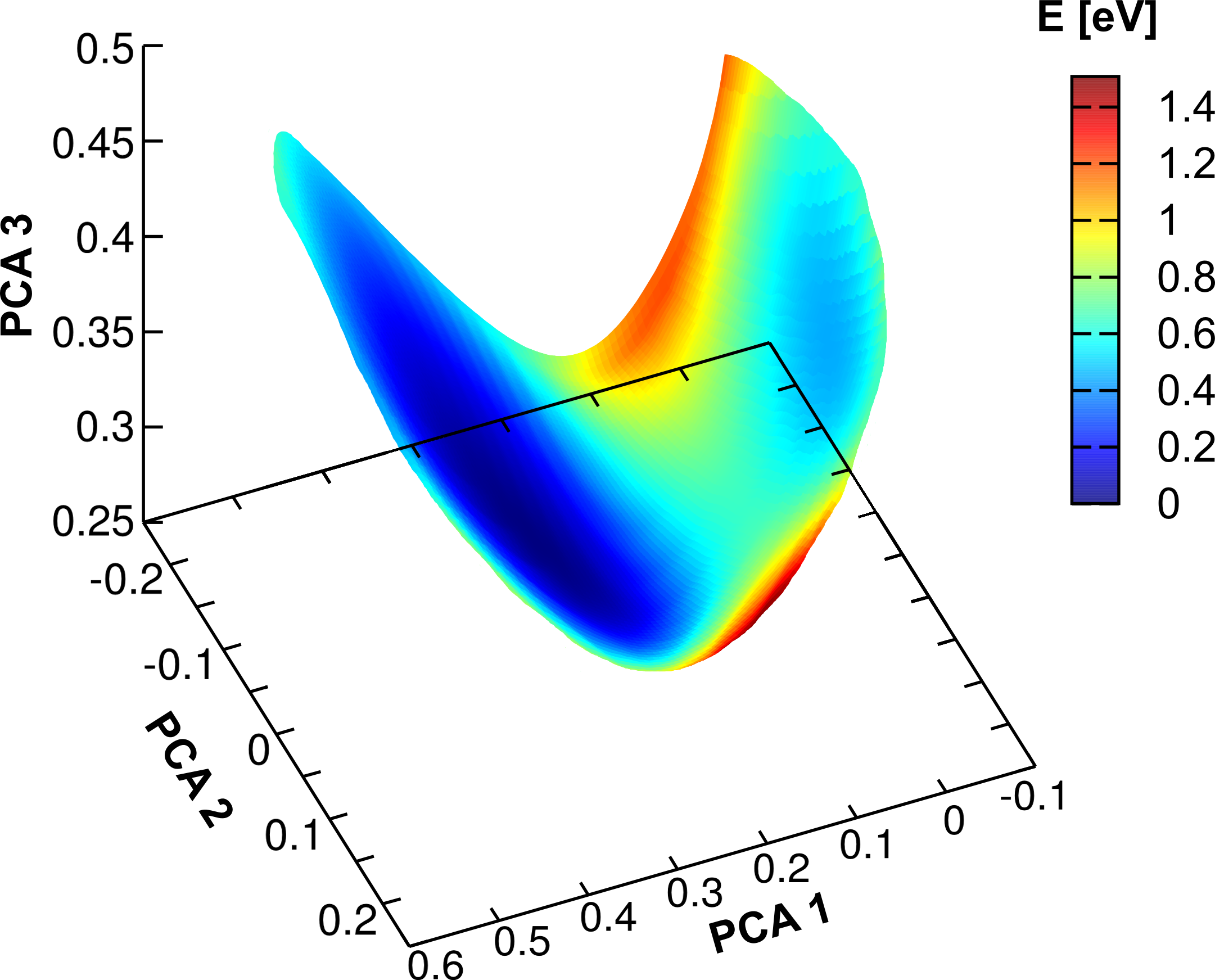} 
\end{center}
\vspace{0.25in}
\hspace*{3in}
  \caption{Projection of the two-dimensional potential energy surface generated by our method onto the first three principal components of the training data set. The \ac{pca} coordinates are given in relative units. Since all three \ac{pca} coordinates contribute significantly to the data set and the \ac{irc} the surface is strongly curved. The depth of the \ac{pes} is given by color and is not related to the spatial dimensions. It shows the double minimum potential structure typical for derivatives of malondialdehyde. This projection is analogous to an angle coordinate projected onto its x and y components, resulting in a circle, on which a color coded potential could be represented.}
  \label{fig:cpot}
\end{figure}

\begin{figure} [ht]
\begin{center}
   \includegraphics[width=0.6\columnwidth,keepaspectratio=true]{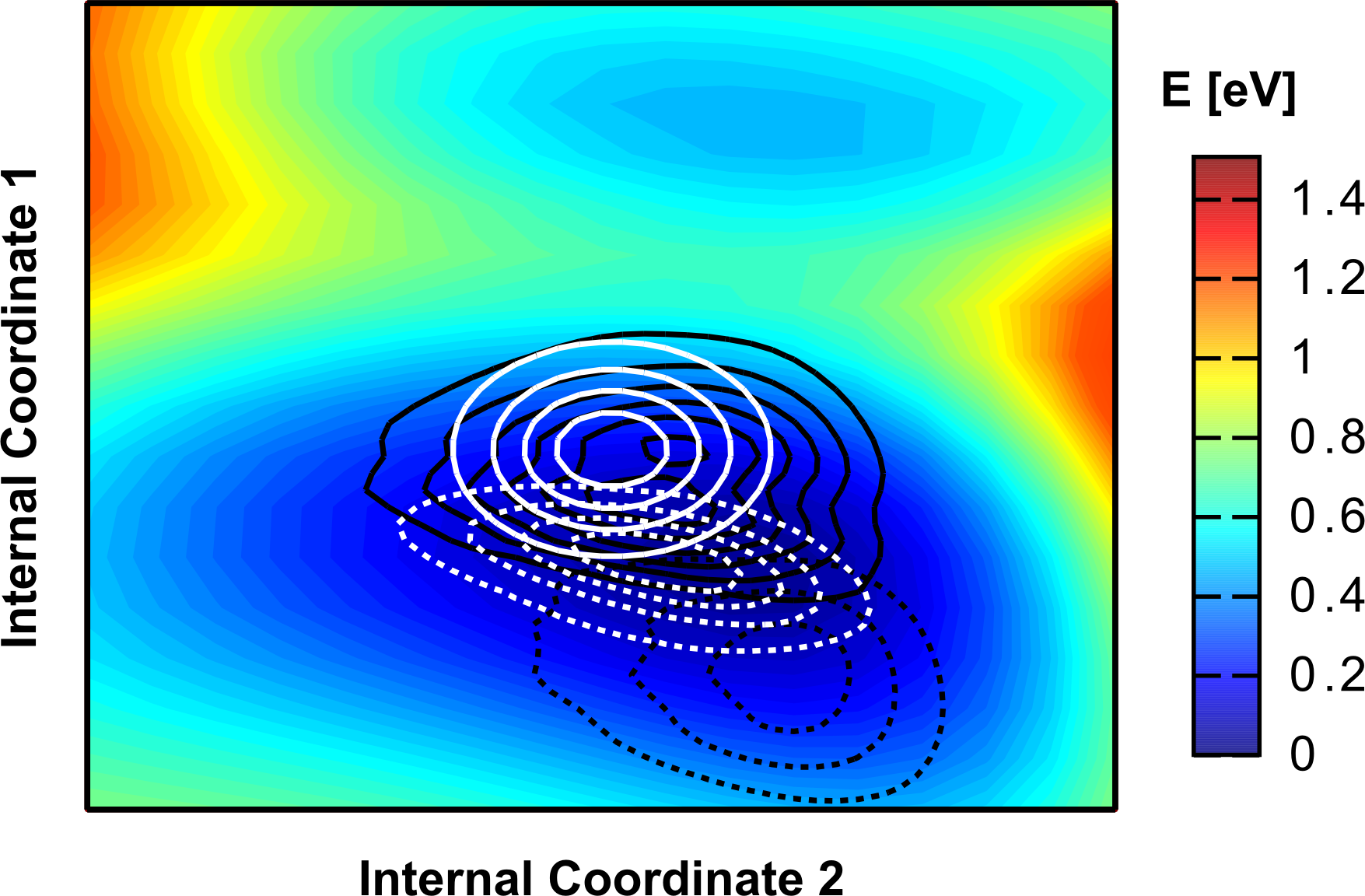} 
\end{center}
\vspace{0.25in}
\hspace*{3in}
  \caption{Wavepacket propagation in two reactive coordinates. The \ac{pes} is shown on a rectangular grid of nonlinear reactive coordinates, constructed using an autoencoder. The wavepacket propagation is shown in snapshots at 0 fs (solid white), 17 fs (dashed white), 31 fs (dashed black) and 68 fs (solid black). The wavepacket travels through the deeper potential minimum and is being reflected.}
  \label{fig:wfpropa}
\end{figure}

Using the autoencoders and eq. \ref{eq:gmat3} \ac{pes} grids of sizes 48$\times$17 and 48$\times$17$\times$24 were constructed as well as a grid of size 100$\times$100 for comparison and visualization at the DFT (M06-2X,6-31G(d)) level of theory. A projection of the 2-dimensional nonlinear subspace \ac{pes} onto the space spanned by the first three principal components is shown in \ref{fig:cpot}. One can see the curvature of the subspace as well as the double minimum structure with the transition state lying inbetween. For this test scenario, a Gaussian wavepacket starting near the transition state was propagated 500 time steps with a step size of about 0.24 fs using the Chebychev propagation scheme \cite{TalEzer1984}. During this time, the wavepacket reached the minimum and was reflected as can be seen for the two-dimensional case in \ref{fig:wfpropa}. Overall, the propagation in the 2- and 3-dimensional cases did not differ from the ones on more traditional grids in terms of stability or conservation of total energy or norm. The wavepacket ran along the reaction path and showed no unphysical behavior and the time scales of the nuclear motion are consistent with the time scales of the keto-enol switching in malondialdehyde \cite{Wolf1998}

\section*{Summary and Discussion}
We presented a method for automated nonlinear dimensionality reduction for reactive molecular systems. It uses an autoencoder trained on trajectory data points to create a low-dimensional representation of reactive coordinates. With this approach, trajectories that describe the space of the studied reaction sufficiently well offer a good basis for \ac{qd} calculations. The autoencoder is able to find a low-dimensional subspace that is close to the configuration space used by the trajectories. We presented all the steps necessary to create the data sets needed, to train the autoencoder and to use it to construct a \ac{qd} grid by projecting points onto the low-dimensional subspace. To use this grid, we gave a short overview of the G-matrix formalism that is useful to construct a kinetic energy operator for this general type of coordinates. 

These steps were then applied to our example system (Z)-hydroxyacryloyl chloride. The high quality of the resulting autoencoder subspaces was demonstrated by their proximity to the \ac{irc}. The \ac{qd} calculations on the resulting grids performed well, without any noticeable unphysical behavior or significant conservation errors in total energy or norm.

There are a few straightforward next steps that could be very beneficial. The first is the application of this method to other systems. There is a huge range of applicable scenarios, but one field that seems especially promising is the laser steering of reactions or optimal control theory as it strongly relies on quantum effects. Classical molecular dynamics can be used to find individual laser pulses that steer reactions for very specific starting conditions and paths, but the pulses differ a lot from each other. However, it is likely that the configuration space used by the individual trajectories is a good predictor for high quality reactive coordinates used in \ac{qd} calculations. Thus, this might be an important step to automate the search for efficiently steering laser pulses in complex molecular systems.

Another step is the improvement of the grid creation strategy. While we are confident that the discovered subspaces are generally of high quality, the projection of a grid in a linear subspace onto the nonlinear subspace might result in very unevenly spaced grid points for highly curved degrees of freedom. However, this problem seems very solvable by developing iterative grid construction algorithms. There are also a number of possible steps to refine the method. For example, it might sometimes be relevant that critical points in the reactive systems are included exactly in the subspace. This might be solved by different kinds of local or global shifts of the created subspace grid in the full configuration space.

Overall, we see our method as very consequential for grid based approaches. It allows the use of the best affordable trajectory method for a system to identify a very representative subspace in which higher level \ac{qd} calculations can be performed. Additionally, the application prospects and the room for simple extensions are very promising.

\begin{acknowledgement}
Financial support by the Deutsche Forschungsgemeinschaft through the SFB749 and the excellence cluster Munich-Centre for Advanced Photonics (MAP) is acknowledged.
\end{acknowledgement}

\begin{suppinfo}
The Supporting Information contains a short derivation of the iterative projection given by eqs. \ref{eq:autoenc3} and \ref{eq:autoenc4} as well as the derivation of the maximum distance between neighbouring grid points as given in eq. \ref{eq:gmat3}.

\end{suppinfo}
\bibliography{litarxiv}

\end{document}


\begin{acronym}
\acrodef{dpm+}[\ce{Ph2CH+}]{diphenylmethyl cations}
\acrodef{dpm.}[\ce{Ph2CH^{\textbullet}}]{diphenylmethyl radicals}
\acrodef{mpm}[\ce{PhCH2}]{phenylmethyl}
\acrodef{dam}[\ce{Ar2CH}]{diarylmethyl}
\acrodef{dpmtpp}[\ce{Ph2CH-PPh3+}]{diphenylmethyltriphenylphosphonium ions}
\acrodef{pca}[PCA]{principal component analysis}
\acrodef{coin}[CoIn]{conical intersection}
\acrodef{irc}[IRC]{intrinsic reaction coordinate}
\acrodef{subc}[SubC]{intrinsic reaction coordinate}
\acrodef{ts}[TS]{transition state}
\acrodef{oniom}[ONIOM]{our own n-layered integrated molecular 
orbital and molecular mechanics}
\acrodef{cas}[CASSCF]{complete active space self consistent field}
\acrodef{dft}[DFT]{density functional theory}
\acrodef{pes}[PES]{potential energy surface}
\acrodef{fc}[FC]{Franck-Condon}
\acrodef{si}[SI]{supporting information}
\acrodef{cqdmd}[cQD/MD]{coupled QD/MD}
\acrodef{qd}[QD]{quantum dynamics}
\acrodef{md}[MD]{molecular dynamics}
\acrodef{tdscf}[TDSCF]{time-dependent self-consistent field}
\acrodef{vrcs}[VRCS]{virtual rigid classical solute}
\end{acronym}

\subsection*{Approximate Projection onto Nonlinear Subspaces}
This projection assumes that the point $\mathbf{y}$ to be projected is close enough to the subspace and the subspace is not pathologically curved. Otherwise the algorithm will only find relative projection optima. It also assumes that the imperfect projection $\widetilde{\mathbf{H}}(\widetilde{\mathbf{F}}())$ does not change drastically with small changes in its argument. In practice, this assumption holds.

We now have a set of non orthonormal basis vectors $\mathbf{a}_j$, comprising a nonorthonormal projection matrix $\mathbf{A}=[\mathbf a_1\ \mathbf a_2\,\cdots\,\mathbf a_N]$. These give us a basis of the tangential space of the nonlinear subspace at the point $\widetilde{\mathbf{H}}\left(\widetilde{\mathbf{y}}'\right)$. If we find a projection onto this tangential space, we can project the distance between the point to be projected $\mathbf{y}$ and the current approximation of its projection $\widetilde{\mathbf{H}}\left(\widetilde{\mathbf{y}}'\right)$ onto it. This projection is approximately the distance vector between the approximate projection and the actual projection. Thus, if we deduct it from $\widetilde{\mathbf{H}}\left(\widetilde{\mathbf{y}}'\right)$ and iteratively do this with an appropriately small step size, we reach the actual projection.

The last thing missing is an expression for a projection using a nonorthonormal basis. To derive this, we start with the most general representation of a vector $\mathbf{x}$ in this basis:
\begin{align}
\mathbf{x}=\sum_i a_i\mathbf{a}_i
\label{eq:si6}
\end{align}
Here $a_i$ is the coefficient of its $i$th basis vector. Those coefficients tell us how the vector $\mathbf{x}$ looks in our basis $\mathbf{a_i}$ and we want to find a vector $\mathbf{b}=[ a_1\  a_2\,\cdots\, a_N]$ composited of these coefficients $a_i$. 

To get there we multiply $\mathbf{x}$ with an $\mathbf{a}_j^T$:
\begin{align}
\mathbf{a}_j^T\mathbf{x}=\sum_i a_i\mathbf{a}_j^T\mathbf{a}_i
\label{eq:si7}
\end{align}
If this is considered to be the $j$th element of a vector, we can write this as
\begin{align}
[ \mathbf{a}_1^T\mathbf{x}\ \  \mathbf{a}_2^T\mathbf{x}\,\ \cdots\, \ \mathbf{a}_N^T\mathbf{x}]^T=\mathbf{x}^T\mathbf{A}=\mathbf{A}^T\mathbf{A}\mathbf{b} \text{ .}
\label{eq:si8}
\end{align}
If we isolate $\mathbf{b}$ we obtain
\begin{align}
\mathbf{b}=\left(\mathbf{A}^T\mathbf{A}\right)^{-1}\mathbf{x}^T\mathbf{A} \text{ .}
\label{eq:si9}
\end{align}

\subsection*{Grid Spacing} 
We begin with the classical expression of the kinetic energy $T$ in terms of the G-matrix $\mathbf{G}$ and the momentum vector $\mathbf{p}$ \cite{Wilsonbook}:
\begin{align}
T=\frac{1}{2}\mathbf{p}^T\mathbf{G}\mathbf{p}
\label{eq:si1}
\end{align}
Now, a relationship between the kinetic energy and the momentum along a single coordinate is needed. In the worst case all of the kinetic energy is contained within a single directional momentum $p_{r,\text{max}}$, which represents the maximum momentum that is classically possible in this direction. This means that the the momentum vector $\mathbf{p}$ has only one non zero entry at index $r$. If we insert this in Eq. \ref{eq:si1}, it becomes
\begin{align}
T=\frac{1}{2}p_{r,\text{max}}^2\mathbf{G}_{rr} \text{ .}
\label{eq:si2}
\end{align}
Since the nyquist frequency determines the upper bound for the spacing $\Delta q_r$ in order to represent a certain momentum via
\begin{align}
\Delta q_r<\frac{\pi}{p_{r,\text{max}}} \text{ ,}
\label{eq:si3}
\end{align}
we can now derive the relationship between the spacing and the kinetic energy along a certain direction as
\begin{align}
\Delta q_r<\pi\sqrt{\frac{\mathbf{G_{rr}}}{2T}} \text{ .}
\label{eq:si4}
\end{align}
In practice the kinetic energy $T$ will almost never be close to completely in coordinates orthogonal to the IRC. Therefore the kinetic energy $T_r$ along the different directions $r$ also differs, resulting in
\begin{align}
 \Delta q_r < \pi \sqrt{\frac{G_{rr}}{2T_r}}\text{ .}
\label{eq:si5}
\end{align}

\bibliography{litarxiv}